# Wohlleben Effect and Emergent π junctions in superconducting Boron doped Diamond thin films

L. Govindaraj[1], S. Arumugam[1]*, R. Thiyagarajan[1,2], Dinesh Kumar[3], M. Kannan[1], Dhrubha Das[3], T. S. Suraj[3,4], V. Sankaranarayanan[4], K. Sethupathi[4], G. Baskaran[5,6,7], Raman Sankar[8,9], M. S. Ramachandra Rao[3]

[1]*Centre for High Pressure Research, School of Physics, Bharathidasan University, Tiruchirappalli 620024, Tamil Nadu, India*

[2]*Institute of Solid State and Materials Physics (IFMP), Dresden University of Technology, D-01069 Dresden, Germany*

[3]*Department of Physics, Nano Functional Materials Technology Centre and Materials Science Research Centre, Indian Institute of Technology Madras, Chennai 600036, India*

[4]*Department of Physics, Low Temperature Physics Laboratory, Indian Institute of Technology, Chennai 600036, India*

[5]*Department of Physics, Indian Institute of Technology, Chennai 6000036;*

[6]*The Institute of Mathematical Sciences, Chennai 600 113, India;*

[7]*Perimeter Institute for Theoretical Physics, Waterloo, Ontario N2L 2Y5, Canada*

[8]*Institute of Physics, Academia Sinica, Nankang, Taipei 11529, Taiwan*

[9]*Centre for Condensed Matter Sciences, National Taiwan University, Taipei 10617, Taiwan*

*****Corresponding author:** e-mail: sarumugam1963@yahoo.com**,** Phone: (O): +91-431-2407118, Cell: +91-95009 10310; Fax No: + 91-431- 2407045, 2407032

**Keywords:** Boron Doped Diamond, Superconductor, paramagnetic Meissner effect, spin glass, two step superconducting transition

## Abstract

Diamond is an excellent band insulator. However, boron (B) doping is known to induce superconductivity. We present two interesting effects in superconducting B doped diamond (BDD) thin films: i) *Wohlleben effect (paramagnetic Meissner effect, PME) and* ii) a low field *spin glass like susceptibility anomaly.* We have performed electrical and magnetic measurements (under pressure in one sample) at dopings (1.4 , 2.6 and 3.6) × $10^{21}$ cm$^{-3}$, in a temperature range 2 - 10 K. PME, a low field anomaly in inhomogeneous superconductors could arise from flux trapping, flux compression, or for non-trivial reason such as emergent Josephson π junctions. Joint occurrence of PME and spin glass type anomalies points to possible emergence of π junctions. BDD is a disordered s-wave superconductor; and π junctions could be produced by spin flip scattering of spin ½ moments when present at weak superconducting regions (Bulaevski *et al.* 1978). A frustrated network of 0 and π junctions will result (Kusmartsev *et al.* 1992) in a distribution of spontaneous equilibrium supercurrents, a phase glass state. Anderson localized spin ½ spinons embedded in a metallic fluid (two fluid model of Bhatt et al.) could create π junction by spin flip scattering. Our findings are consistent with presence of π junctions, invoked to explain their (Bhattacharyya et al.) observation of certain resistance anomaly in BDD.

## I. INTRODUCTION

Diamond, a wide gap band insulator, is unique in many respects, in the form of single crystals, thin films or nanodiamonds. Thin films of diamonds are used in electronics industry as microchip substrates, high-efficiency electron emitters, photo detectors, transistors, electrodes in electrochemistry etc. [1]. Boron doped diamond (BDD) was known to exhibit an insulator to metal transition [2] at a critical doping. Boron, an acceptor has one less electron compared to carbon. In 2004, Ekimov and collaborators [3] surprised the scientific community, by their discovery of insulator to superconductor transition in diamond as a function of B doping. Insulator to type II-superconductor transition occurs at a critical B doping ~ $0.3 \times 10^{21}$ cm$^{-3}$. Superconductivity survives for a range of doping, with $T_c$ ranging from 50 mK (close to critical doping) to 11.4 K for higher dopings (~ $8.4 \times 10^{21}$ cm$^{-3}$) [4,5].

Following the pioneering discovery of Ekimov *et al.* [3], a wealth of experimental and theoretical activities followed [4-18]. Phenomenological superconducting parameters, vortex structures and superconducting gap structures have been measured experimentally. Spatial disorder, arising random B substitution, inhomogeneity, preparation methods etc., create superconductivity with widely differing $T_c$'s. An onset of 25 K and a zero resistance state at 10 K has been also reported [11]. Even higher $T_c$ ~ 55 K has been reported in B doped Q carbon, proposed as a diamond rich carbon system [12].

To gain more insight into nature of superconductivity in BDD thin films (prepared by CVD method) we studied transport and magnetic properties of three different BDD thin films (1.4, 2.6 and 3.6) × $10^{21}$ cm$^{-3}$ and find two magnetic effects at low magnetic fields (~ few tens of Oe), coexisting with superconductivity: i) Wohlleben effect (paramagnetic Meissner effect, PME) [19-24] and ii) pressure induced spin glass like anomaly in low field susceptibility.

PME, widely discussed in cuprates, has also been observed in other conventional and unconventional superconductors. PME exhibits significantly different behaviour of zero field cooled (ZFC) and (small) field cooled (FC) magnetic susceptibility. Ekimov *et al.,* [3] and later works observed significant difference in ZFC and FC susceptibility below $T_c$. Attention was not paid to this and it was generally attributed to flux trapping. But our finding of simultaneous presence of a pronounced PME and a spin glass type anomaly becomes interesting, as it points to an intrinsic and microscopic origin - emergent Josephson $\pi$ junctions.

We have also studied pressure effect on PME in one of our sample (doping ~ 2.6 x $10^{21}$ cm$^{-3}$) under various magnetic fields, and found a new result, indicative of presence of spin glass (phase glass) type order ($T_{SG} < T_c$). In one of our samples (doping concentration: 3.6 x $10^{21}$ cm$^{-3}$), we also observe a two-step superconductivity transition. Two step transitions have been observed and discussed earlier experiments [25,26] in BDD.

PME arises for a variety of reasons: flux trapping, flux compression, presence of Josephson coupled $\pi$ junctions, PT violating spontaneous supercurrents etc. $\pi$ junctions could arise because of unconventional order parameter (d-wave symmetry in cuprates, as suggested by Sigrist and Rice [22]). But BDD is a disordered s-wave spin singlet superconductor. Localized spin moments, when present at weak links in an s-wave superconductor could cause $\pi$ phase shifts, via spin-flip scattering - a mechanism suggested by Bulaevski, Kuzii and Sobyanin [27]. Further, odd number of $\pi$ junctions will generate spontaneous ground state supercurrents. Kuzmartsev [28] has suggested that a random collection of 0 and $\pi$ junctions, a frustrated system, could create a glassy equilibrium supercurrent phase, called orbital glass or chiral glass. Independent of PME, phase glass in highly disordered superconductors (for zero or small fields) and field induced vortex glass states have been discussed in the past[29,30]. Experimentally a high field vortex glass phase, close to $H_{c2}$ has been reported in BDD [31]. What we are proposing for our present result is an *intrinsic phase glass phase*, whose presence is revealed via low magnetic field response, in the absence of a high uniform external magnetic field.

In an important work Bhattacharyya and collaborators [32] have observed certain negative resistance anomaly in BDD and attribute it to possible presence of $\pi$ junctions. They invoke non s-wave (p-wave) order parameter as possible origin of $\pi$ junctions. As we mentioned earlier, BDD is a disordered spin singlet s-wave superconductor; s-wave symmetry is robust and protected by Anderson theorem [33]. So the $\pi$ junctions proposed by Bhattacharyya *et al.,* is likely to arise from spin-flip scattering mechanism we are invoking in this article, rather than from an unconventional order parameter.

In doped semiconductors disorder and electron correlations are known to play important role close to the metal to insulator transition region in non-magnetic semiconductors. Certain generic and key consequences have been suggested: i) Coexistence of conducting fluid and spatially (Anderson) localized dopant spin-moments on the conducting side (Bhatt, Paalanen and Sachdev, BPS Model) [34] and ii) valence bond glass (VBG) phase on the insulating side [35]. We suggest that these localized spin ½ moments play a crucial role in creating $\pi$ junctions in our BDD samples, via spin flip scattering mechanism [27]. This explains, in a unified fashion, simultaneous presence of Wholleben effect and spin glass like anomaly.

We briefly point out how our results support i) Bhatt-Paalanen-Sachdev model [19], ii) also an indirect support to Bhatt-Lee model [20] of valence bond glass and iii) a model of one of us [17] for RVB theory of superconductivity in impurity band Mott insulator, via self doping in BDD. Here, increasing concentration of B atom is equivalent to increase of effective internal chemical pressure on the acceptor impurity band; certain organic Mott insulators undergo transition to a superconducting state, via self doping [36], beyond a critical external pressure.

## II. PREPARATION AND EXPERIMENTAL METHODS

As Boron has a smaller atomic radius and right quantum chemistry, boron doping in diamond is achieved in the Chemical Vapour Deposition (CVD) method [37]. We prepared thin film samples with three different boron dopings by the Hot Filament Chemical Vapour Deposition (HFCVD) technique. Detailed method of preparation is already discussed by Mahmoud Abdel-Hafiez, *et al.* [38]. To estimate B doping density, Raman spectra on each sample were taken using Renishaw InVia Raman microscope system in the wave number range of 200 to 700 cm$^{-1}$ with excitation laser source of 532 nm. Electrical resistivity measurements were carried out using Physical Property Measurement System (PPMS), DynaCool (Quantum design, USA) by the conventional four-probe method. Initially, four contacts on sample were made by high-quality silver paste with a copper wire of Φ 0.05 mm. Temperature dependence of electrical resistivity measurements [$\rho(T)$] were performed under various magnetic fields (up to 7 T) and temperature interval of 2 to 10 K, with steps of 0.01 K. Magnetization measurements were performed using PPMS-Vibrating Sample Magnetometer module (PPMS-VSM, Quantum Design, USA). A clamp type miniature hydrostatic pressure cell (MCell from Easylab, UK) made of nonmagnetic Cu-Be alloy was used for the high-pressure magnetization measurements up to ~ 1 GPa on $2.6 \times 10^{21}$ cm$^{-3}$ sample. The mixture of flourinert #70 and flourinert #77 was used as a pressure-transmitting medium. The value of pressure was estimated from the shift of the superconducting critical temperature with applied pressure of pure Sn, earlier to each temperature dependence of magnetization [$M(T)$] measurements [39]. The $M(T)$ were recorded during Zero Field Cooling (ZFC) mode in the temperature range of 2 to 10 K under various magnetic field strengths in steps of 10 Oe, till the observed-transitions survived, for several fixed hydrostatic pressures. The temperature is swept at the rate of 0.05 K/min and then the sample has been stabilized at each temperature for 1 minute. Previously, the pressure cell has been kept for 4 h at 10 K and 2 h at 2 K in order to cool the sample homogeneously for required temperatures within the pressure cell. As typical measurement setting for VSM with pressure cell is 20 Hz, it took approximately 16 h for each set of data. The temperature error is less than 0.05 K, which is negligible. The same set of measurements has been repeated for two times to confirm the reproducibility of the transitions both at ambient and high pressures.

## III. EXPERIMENTAL RESULTS

### III.a. Raman Measurements and B doping Estimation

There are few ways of determining B concentration in BDD, including a non-destructive Raman measurement. Bernard *et al.*, suggested [40] an empirical relationship between position of a Raman peak (W, wave number in cm$^{-1}$) and boron concentration in boron doped diamond as $\approx$ $8.44 \times 10^{30} \times e^{-0.048 W}$ cm$^{-3}$. Fig. 1 shows Raman spectra of our three samples around the 500 cm$^{-1}$ peak with its corresponding Lorentzian fittings. The shifting of peaks towards higher wave number has been ascribed to boron content of the films. From our Raman data, we estimate the concentrations to be $1.4 \times 10^{21}$, $2.6 \times 10^{21}$ and $3.6 \times 10^{21}$ cm$^{-3}$. In what follows we call them sample A, sample B and sample C respectively.

### III.b. Electrical Resistivity Measurements

Fig. 2(a) shows $\rho(T)$ of all the three samples, for zero magnetic field, by reducing the temperature from 10 K. Samples A, B and C exhibited slightly different superconducting transitions ($T_c^{onset}$) at 5.6 K, 6.1 K and 6.7 K respectively. In sample C, an additional sharp drop has been observed in $\rho(T)$ at 3.9 K. Such a step has been seen in earlier experiments and discussed [25,26]. Fig. 2(b-d) shows $\rho(T)$ of the samples A, B and C respectively under various magnetic fields up to 7 T. Increasing magnetic field shifts $T_c$ to lower values, and reduces $T_c^{onset}$ correspondingly. Superconducting $T_c$'s of samples A, B and C as the function of magnetic field are shown in Fig. 3(a-c). $T_c^{onset}$ of samples A, B and C are reduced at the rate of - 0.76 K/T, - 0.4 K/T and - 0.71 K/T respectively. Location of second step in resistivity of sample C is also reduced by magnetic field at the rate of - 0.76 K/T.

### III.c. Magnetization Measurements

Fig. 4(a-c) shows $M(T)$ of samples A, B and C respectively under various magnetic fields. While reducing the temperature from 10 to ~ 6 K at ambient pressure with 10 Oe of magnetic field, the magnetization appears with low positive values which suggest the existence of paramagnetic behaviour. Further reducing the temperature, sample A and B show a paramagnetic peak at 3.9 and 5.4 K for 10 Oe. This is a clear case of PME [41]. By increasing the magnetic field, diamagnetism of sample A and B vanished at 250 and 100 Oe respectively. Sample C under 10 Oe exhibits a weaker PME and its corresponding superconductivity at 6 K. By increasing the magnetic field, diamagnetism of sample C vanishes at 1200 Oe.

### III.d. Pressure effects on magnetization

In order to study the pressure effect on BDD, $M(T)$ measurements were carried out on sample under various magnetic fields for fixed hydrostatic pressures of 0.0, 0.05, 0.5 and 0.9 GPa, shown in Fig. 5(a-d) respectively. The y-axes are given in arbitrary scales for the sake of clarity. The data of M for these Figures in CGS units are given as Supplementary [Fig. S1(a-d)]. Under ambient pressure and a magnetic field of 10 Oe, as expected, $T_c$ has been observed at 5.44 K. While increasing the magnetic field for various pressures, magnitude of magnetization increases and $T_c$ shifts to lower temperature (< 5.0 K). As we are dealing with granular material, suppression of superconducting $T_c$ could arise from different reasons. Interestingly, above ambient pressure, we observe a new small kink below $T_c$, and we term it as spin glass transition ($T_{SG}$) [30,42], spin glasses are known to exhibit a similar kink in low field susceptibility at the spin glass transition point.

Estimated $T_c$ and $T_{SG}$ from our data, Fig. 5 (a-d), are presented as a phase diagram in Fig. 6. At the pressures of 0.0, 0.05 and 0.5 GPa, superconductivity ceases at 80, 50 and 20 Oe respectively, and there is no trace for superconductivity at and above 0.9 GPa [Fig. 5(d)]. At 0.05, 0.5 and 0.9 GPa, $T_{SG}$ has been observed at 3.5, 3.3 and 3.2 K respectively. Upon increasing the magnetic field and at 0.05, 0.5 and 0.9 GPa, this transition has also been found to be suppressed at 110, 140 and 150 Oe respectively.

## IV. DISCUSSION

### IV.a. Paramagnetic Meissner Effect

PME is an unusual response of a field cooled (FC) superconductor at very low magnetic fields (low compared to $H_{c1}$), which is very different from zero field cooled (ZFC). External magnetic field applied to a flux free superconductor will induce screening supercurrents that opposes applied magnetic field - this is the standard diamagnetic Meissner response. If supercurrents preexist in the sample, applied magnetic field will reorganize them. Pre-existing supercurrents could be of non-equilibrium origin, such as metastable trapped magnetic fluxes in a granular medium. However, if intergranular π junctions are present, spontaneous equilibrium supercurrents could emerge. Closed loops containing *odd number of π junctions* will lead to spontaneous equilibrium supercurrents.

In the case of PME, paramagnetic response (2$^{nd}$ term of equation 1) completely mask the diamagnetic Meissner response (first term):

$$\chi = -\chi_0 + \frac{M_0}{(H+H_0)^\alpha} \qquad (1)$$

Here $H_0$ ~ 1 Oe, $M_0$ ~ 1 Gauss/4π and α ~ 1. PME or Wholleben effect was extensively investigated for high $T_c$ cuprates. A mechanism suggested by Sigrist and Rice invoked [22] π junctions in granular high $T_c$ materials, arising from the unconventional d-wave (sign changing under π/2 rotation) order parameter. For s-wave superconductors, π junction arising from spin flip scattering via paramagnetic spin has been also proposed [27].

There is evidence experimentally and theoretically that disordered BDD is an s-wave superconductor. Pairing takes place in spin singlet channel, between non-Bloch single particle eigen states related by time reversal symmetry and their s-like in-phase combination, in the sense of Anderson [33] for dirty s-wave superconductors. Absence of periodicity of dopant B atoms makes d-wave order parameter (which demand presence of square or cubic symmetry) energetically unfavourable. Our BDD thin film samples are homogenous at the scale of size of quantized vortices (London penetration length ~ 150 nm). Thus trapping and flux compression seem unlikely. Induced $\pi$ junctions by spin flip scattering of localized paramagnetic moments is an interesting possibility.

**IV.b. Origin of Localized Paramagnetic Moments**

For very low doping in BDD, well below the critical doping of metal insulator transition, spatially isolated neutral B dopants, a neutral acceptor ($B^0$) carries spin-half moment of a hole. As doping density increases near neighbour acceptor state wave functions start to overlap. Resulting kinetic (super) exchange leads to spin singlet formation of B moment pairs. In the theory of Bhatt and Lee [35], singlet bond formation happens in a hierarchical (weak to strong) fashion and leads to a valence bond glass (VBG) phase. VBG is a non-trivial phase that also supports local valence bond resonances, or RVB puddles. Consequently this state exhibits a remarkable fractional power law divergence of paramagnetic susceptibility, $\chi \sim 1/T^{(1-\alpha)}$ (with positive $\alpha$), rather than an exponential decrease, ¯at low temperatures, expected from a valence bond solid with a spin gap. Thus existence of isolated or weakly coupled paramagnetic spin of dopants is obvious on the insulating side.

What happens on the metallic side of the metal insulator transition point? Based on phenomenology and microscopic considerations, Bhatt-Paalanen-Sachdev (BPS) presented a two fluid model for the conducting side of the metal to insulator transition point: a conducting electron fluid and a collection of localized spins weakly coupled to it. When we use BPS model for BDD, in the superconducting phase we expect a finite density of localized spins which are weakly coupled to the disordered superconducting condensate. In the RVB terminology this could be viewed as spatially separated Anderson localized neutral spin-1/2 spinons present in the ground state.

**IV.c. Emergent $\pi$ Junctions from Spin Flip Scattering**

A paramagnetic spin-half moment acts like a pair breaker for a singlet superconductor. However, a spin-half moment present in the weak link region between two superconductors can cause a non-trivial additional effect, as predicted by Bulaevski et al [27]. It can cause a phase change $\pi$ between phases of the superconducting order parameter across the junction. Sign of energy of Josephson coupling effectively changes.

We propose that local spin moments present in our system, as suggested by the two fluid model of BPS can cause the required spin flip scattering and create required π junctions between superconducting granular regions. Bhattacharya and collaborators [32] have also suggested presence of π Josephson junction in their BDD from their observation of a negative resistance anomaly. We believe that the mechanism for emergence of π junctions we are invoking works in their case as well.

**IV.d. Phase (Chiral/Orbital) Glass as origin Spin Glass like Anomaly**

One of our samples shows a kink in susceptibility as a function of temperature, in the presence of small applied pressure. This is superimposed on the paramagnetic Meissner signal, below the superconducting $T_c$. Kink in low field susceptibility is a characteristic of transition to a spin glass phase in disordered system systems [30,42,43]. Further, this spin glass type anomaly shifts to lower temperatures as pressure is increased.

In our proposed picture a network of randomly located π junctions are present in our system. Following Kursmastev this can be viewed as an effective random lattice of 0 and π Josephson junctions [28]. This frustrated phase coupled system can exhibit orbital glass (orbital supercurrent), also called chiral glass order. This glass phase will carry equilibrium supercurrents in a random and glassy fashion.

It should be noted that even in the absence of π junctions, a randomly coupled normal Josephson junctions could exhibit glassy order. Disorder even without frustration [29], for example a diluted ferromagnetic XY model, could create spontaneous vortex antivortex pairs in 2D and vortex loops in 3D in the ground state. In the presence of large external magnetic fields Abrikosov vortex lattice become glassy [30]. Glassy phase reported in BDD at high fields close to $H_{c2}$ is likely to be this vortex glass phase.

## V. CONCLUSION

In the present article, we have presented magnetization and resistivity measurements on thin films of BDD at low magnetic fields. We find: *i)* paramagnetic Meissner (Wohlleben) effect in $M(T)$ and *ii)* pressure-induced spin glass type anomaly below superconducting $T_c$ in one sample. We have suggested that two effects together imply emergence of pi junctions in the bulk of our films of BDD, from structural inhomogeneity and spin flip scattering mechanism. Our finding provides some clue to microscopic origin of superconductivity, either directly or indirectly.

One of us proposed RVB theory of impurity band Mott insulator to supconductor transition in B doped diamond, as a function of B doping. This superconducting state is a natural evolution of Bhatt-Lee's insulating valence bond glass state to a conducting side, across the metal insulator transition point. In the RVB theory of superconductivity, a small and equal density of doublons and holons ($B^+$ and $B^-$) get spontaneously created (self doping) at the impurity band Motti insulator to metal (first order) transition. As the system is a disordered and correlated state, superconducting state accommodates localized spin moments, as in Bhatt-Sachdev-Paalanen model. In the language of disordered RVB theory they correspond to Anderson localized spinons, which are neutral spin ½ quasi particles.

From physics point of view, our finding of PME and a spin glass-*like* magnetic anomaly underscores possible importance of electron correlation effects in impurity bands, in establishing the needed coexistence of superconductivity with localized spin moments. To substantiate our proposal It will be important to establish modified Curie susceptibility $\chi \sim 1/T^{(1-\alpha)}$, (Bhatt and Lee, predicted for P doped Si) experimentally. A real space imaging and study of tunnelling characteristics of localized spinons, appearing in a metallic region will also give further support to the prediction of BPL Unlike the case of P doped Si, BDD offers a comfortable high energy and temperature scales, as well as materials advantage to perform suggested studies. We encourage non-linear susceptibility and magnetization relaxation studies to confirm our finding of a phase glass transition.

## ACKNOWLEDGEMENTS

The author S.A. acknowledges the funding agencies of DST (SERB, FIST, PURSE), RUSA, BRNS and UGC-DAE Consortium for Scientific Research (Indore) for the financial support. G.B. wishes to thank SERB (Department of Science and Technology, India) for a Distinguished

Fellowship; and Indian institute of Technology, Chennai for a Distinguished Professorship. Work done at Perimeter Institute of Theoretical Physics, Waterloo, Canada was supported via a Distinguished Visiting Research Chair. MSR would like to acknowledge the grant received from DST Nanomission (SR/NM/NAT/02-2005) that led to the establishment of Nano Functional Materials Technology Centre (NFMTC) that facilitated diamond growth and the required characterization tools. The author R.T acknowledges the Impuls- und Vernetzungsfonds of the Helmholtz Gemeinschaft e. V. via Helmholtz Exzellenznetzwerk ExNet-0027-Phase 2-3.## References

1. C. J. H. Wort, and R. S. Balmer, Diamond as an electronic material, Mater. Today. **11**, 22–28 (2008). https://doi.org/10.1016/S1369-7021(07)70349-8

2. J.P. Lagrange, A. Deneuville, and E. Gheeraert, Activation energy in low compensated homoepitaxial boron-doped diamond films, Diamond Relat. Mater. **7,** 1390-1393 (1998), https://doi.org/10.1016/S0925-9635(98)00225-8

3. E. A. Ekimov, V. A. Sidorov, E. D. Bauer, N. N.Mel'nik, N. J. Curro, J. D. Thompson, & S. M. Stishov, Superconductivity in diamond. Nature. **428,** 542–545 (2004). https://doi.org/10.1038/nature02449

4. E. Bustarret, J. Kačmarčik, C. Marcenat, E. Gheeraert, C. Cytermann, J. Marcus, and T. Klein, Dependence of the Superconducting Transition Temperature on the Doping Level in Single-Crystalline Diamond Films. Phys. Rev. Lett. **93,** 237005 (2004). https://doi.org/10.1103/PhysRevLett.93.237005

5. Y. Takano, T. Takenouchi, S. Ishii, S. Ueda, T. Okutsu, I. Sakaguchi, H. Umezawa, H. Kawarada, and M. Tachiki, Superconducting properties of homoepitaxial CVD diamond. Diamond Relat. Mater. **16,** 911 (2007). https://doi.org/10.1016/j.diamond.2007.01.027

6. Yoshihiko Takano, Masanori Nagao, Isao. Sakaguchi, Minoru Tachiki, Takeshi Hatano, Kensaku Kobayashi, Hitoshi Umezawa, and Hiroshi Kawarada, Superconductivity in diamond thin films well above liquid helium temperature. Appl. Phys. Lett. **85,** 2851–2853(2004). https://doi.org/10.1063/1.180238911

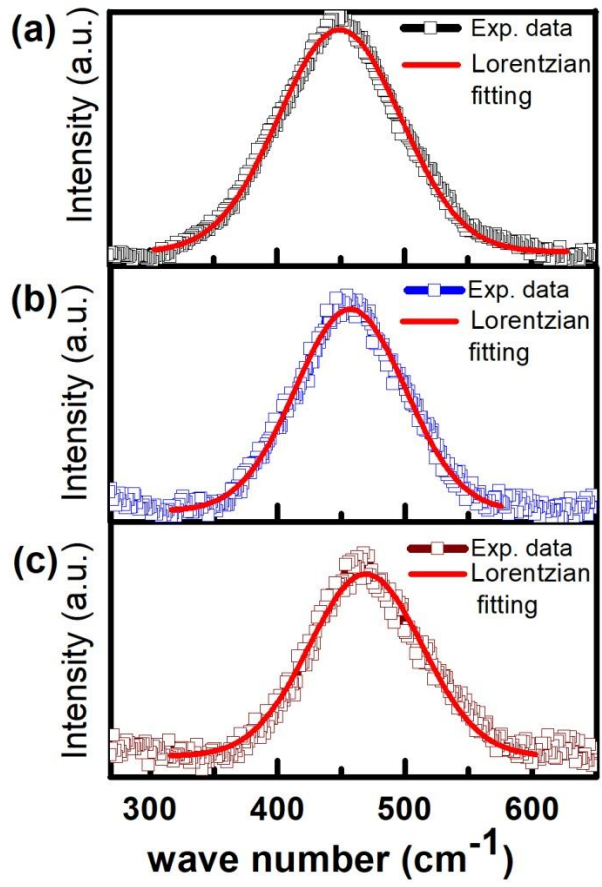

Figure 1: Raman spectra of boron doped diamond thin films. (a) Sample A: $1.4 \times 10^{21}$ cm$^{-3}$; (b) Sample B: $2.6 \times 10^{21}$ cm$^{-3}$; (c) Sample C: $3.6 \times 10^{21}$ cm$^{-3}$

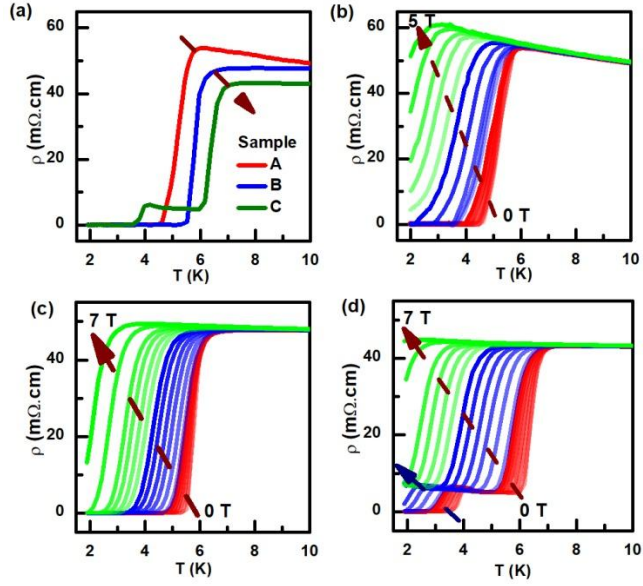

Figure 2: Temperature dependence of resistivity of boron doped diamond thin films. (a) under magnetic field of 0 T; b) sample A: $1.4 \times 10^{21}$ cm$^{-3}$, (c) sample B: $2.6 \times 10^{21}$ cm$^{-3}$ (d) sample C: $3.6 \times 10^{21}$ cm$^{-3}$ under various magnetic fields

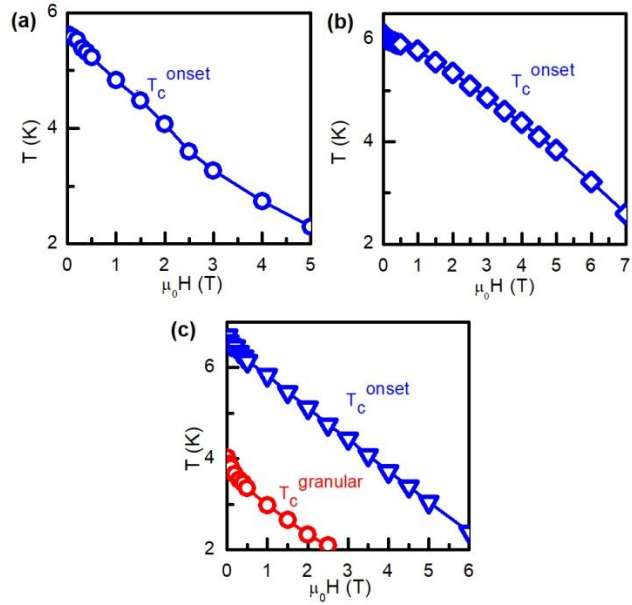

Figure 3: $T_c$ as the function of magnetic field of boron doped diamond thin films. (a) sample A: $1.4 \times 10^{21}$ cm$^{-3}$, (b) sample B: $2.6 \times 10^{21}$ cm$^{-3}$; (c) $T_c$ (blue rectangles) and $T_c^{granular}$ (red circles) of sample C: $3.6 \times 10^{21}$ cm$^{-3}$.

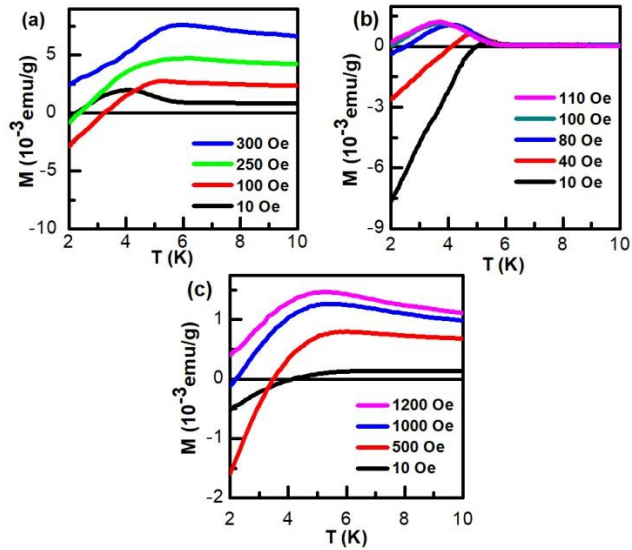

Figure 4 Temperature dependence of magnetization of boron doped diamond thin films. (a) sample A: $1.4 \times 10^{21}$ cm$^{-3}$, (b) sample B: $2.6 \times 10^{21}$ cm$^{-3}$, (c) sample C: $3.6 \times 10^{21}$ cm$^{-3}$ under various magnetic fields

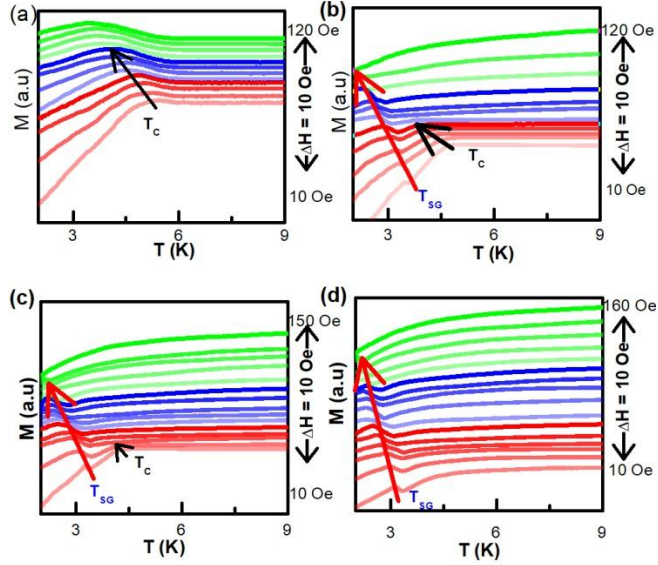

Figure 5: Temperature dependence of magnetization (in arbitrary units) of boron doped diamond thin film (sample B: $2.6 \times 10^{21}$ cm$^{-3}$) under various magnetic fields for fixed hydrostatic pressures. (a) 0 GPa, (b) 0.05 GPa, (c) 0.50 GPa and (d) 0.90 GPa.

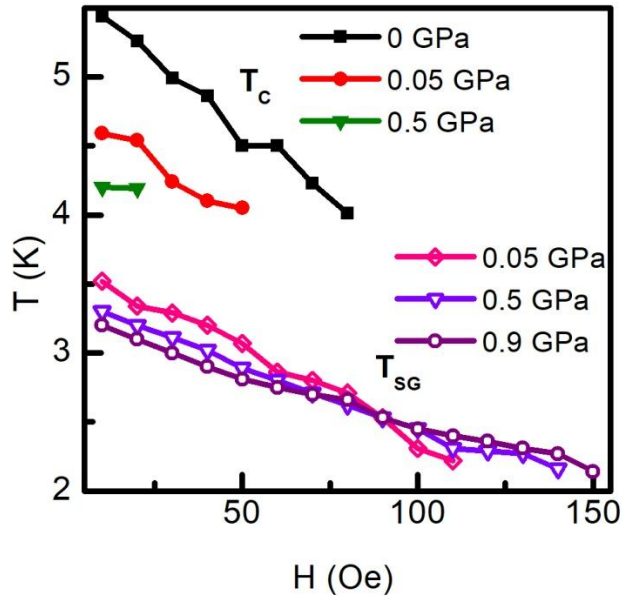

Figure 6: Transition temperatures. $T_c$ and $T_{SG}$ as a function of magnetic field under various fixed pressures on boron doped diamond thin film (sample B: $2.6 \times 10^{21}$ cm$^{-3}$).

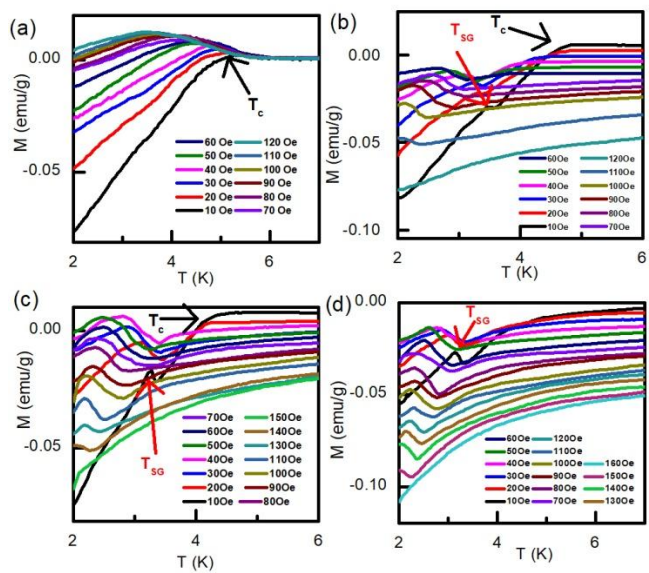

Supplementary Figure S1: Temperature dependence of magnetization (in CGS units) of boron doped diamond thin film (sample B: $2.6 \times 10^{21}$ cm$^{-3}$) under various magnetic fields for fixed hydrostatic pressures. (a) 0 GPa, (b) 0.05 GPa, (c) 0.50 GPa and (d) 0.90 GPa.